\RequirePackage{xcolor}

\documentclass[journal,twoside,web]{ieeecolor}
\usepackage{generic}
\usepackage{cite}
\usepackage{amsmath,amssymb,amsfonts}
\usepackage{algorithmic}
\usepackage{graphicx}
\usepackage{textcomp}
\usepackage{booktabs}
\usepackage{multirow}
\usepackage{url}
\usepackage[hidelinks]{hyperref}
\usepackage{pifont}
\newcommand{\cmark}{\ding{51}}
\newcommand{\xmark}{\ding{55}}

\def\BibTeX{{\rm B\kern-.05em{\sc i\kern-.025em b}\kern-.08em
    T\kern-.1667em\lower.7ex\hbox{E}\kern-.125emX}}
\begin{document}
\title{Unsupervised Susceptibility Distortion Correction of EPI without Calibration Scans via Image Translation-Based Registration}
\author{Wooseung Kim and Sung-Hong Park
\thanks{This work was supported by the National Research Foundation of Korea under Grants RS-2024-00345512 and RS-2026-25527795, and by the Korea Health Industry Development Institute under Grant RS-2024-00354741 (Corresponding author: Sung-Hong Park).}
\thanks{The authors are with the Department of Bio and Brain Engineering, Korea Advanced Institute of Science and Technology, Daejeon, Republic of Korea (e-mail: wooseung.kim@kaist.ac.kr; sunghongpark@kaist.ac.kr).}
}

\maketitle

\begin{abstract}
Functional magnetic resonance imaging (fMRI) utilizes echo-planar imaging (EPI) to capture blood-oxygen-level-dependent (BOLD) signals with high temporal resolution. However, EPI is inherently sensitive to magnetic field inhomogeneities, resulting in susceptibility-induced geometric distortions along the phase-encoding (PE) direction. To correct these distortions, conventional approaches rely on additional calibration scans, such as field maps or reverse PE acquisitions, which are not always available in practice. To overcome this limitation, we propose SACRED, a calibration scan-free susceptibility distortion correction framework that corrects geometric distortions via image translation-based registration using only a routinely acquired anatomical T1-weighted (T1w) image and a unidirectional PE BOLD image. SACRED employs an invertible neural network as the image translation backbone to bridge the contrast gap between BOLD and T1w images while enforcing structural consistency through a modality independent neighborhood descriptor. This design enables the use of a mono-contrast similarity objective to train the registration network in an unsupervised manner without requiring distortion-corrected BOLD images. In addition, we incorporate test-time adaptation (TTA) to further enhance performance on out-of-distribution (OOD) data at inference time. SACRED was evaluated on one in-distribution (ID) dataset and two OOD datasets, and was compared with representative fMRI distortion correction methods. The results demonstrate that SACRED significantly outperforms competing methods on both ID and OOD datasets, exhibiting robustness to scanner and population shifts, partly enabled by TTA. The code will be made publicly available upon acceptance.
\end{abstract}


\section{Introduction}
\label{sec:introduction}
\IEEEPARstart{F}{unctional} magnetic resonance imaging (fMRI) leverages blood-oxygen-level-dependent (BOLD) signals to noninvasively measure neural activity with high temporal resolution~\cite{ogawa_brain_1990}. This temporal resolution is primarily enabled by echo-planar imaging (EPI)~\cite{stehling_echo-planar_1991}, which acquires an entire two-dimensional (or three-dimensional) k-space following a single radiofrequency excitation, enabling rapid whole-brain coverage. However, EPI is inherently sensitive to magnetic field inhomogeneities, largely driven by susceptibility effects, resulting in geometric distortions that predominantly occur along the phase-encoding (PE) direction~\cite{ludeke_susceptibility_1985}. These distortions degrade spatial fidelity and hinder accurate interpretation in downstream analyses~\cite{togo_effects_2017, li_preprocessing_2025}.

To correct susceptibility-induced geometric distortions, conventional EPI distortion-correction methods (e.g., FUGUE and TOPUP from the FSL toolbox~\cite{smith_advances_2004}) typically rely on additional calibration scans, such as a B0 field map or reverse PE images~\cite{jezzard_correction_1995, andersson_how_2003, hwang2023distortion}. These calibration scans enable the estimation of a voxel displacement map (VDM), which is subsequently used to unwarp distorted EPI images. However, acquiring such scans is often constrained by protocol design and scan time. Moreover, many publicly available fMRI datasets do not include calibration scans, limiting the applicability of these approaches~\cite{montez_using_2023}. Even when calibration scans are available, subject motion during the acquisition of these scans can compromise the accuracy of VDM estimation, potentially resulting in residual geometric distortion after correction~\cite{ANDERSSON20161063}.

To address these limitations, several calibration scan-free strategies have been explored. Early studies investigated optimization-based non-linear registration approaches that maximize mutual information between distorted BOLD images and anatomical images~\cite{studholme_accurate_2000, gholipour_cross-validation_2008}. However, the substantial contrast gap between BOLD and anatomical images makes such optimization challenging. To mitigate this issue, subsequent studies focused on image-processing-based approaches to reduce the contrast mismatch between anatomical and BOLD images, thereby facilitating non-linear registration~\cite{kybic2000unwarping, montez_using_2023, esteban_fmriprep_2019}. Nevertheless, these conventional analytical approaches rely on predefined mathematical priors, which limit their ability to model the complex relationships underlying BOLD image contrast. More recently, neural network-based distortion correction methods have been proposed~\cite{yu2023distortion, jimeno2024gdcnet}. Despite their greater representational flexibility, these learning-based approaches can be sensitive to distribution shifts arising from variations in scanner hardware, acquisition protocols, or subject populations, often resulting in degraded performance on out-of-distribution (OOD) datasets~\cite{su2024navigating}.

To overcome the above limitations, we propose an unsupervised framework termed SACRED (Susceptibility Artifact Correction without Reverse phase-Encoding using Deep learning), which leverages image translation-based deformable registration. Building upon our preliminary work~\cite{Kim2026_42bfb763}, we employ an invertible neural network (INN)~\cite{dinh2014nice, dinh_density_2017, kingma_glow_2018} as the image translation backbone to synthesize an undistorted pseudo BOLD reference from the anatomical T1-weighted (T1w) image while preserving structural consistency. This design enables the effective use of standard mono-contrast similarity metrics for training the registration network. In addition, we incorporate a modality independent neighborhood descriptor with self-similarity context (MIND-SSC)~\cite{heinrich2013towards} exclusively as a structure-preserving constraint within the image translation module. Furthermore, to enhance robustness to distribution shifts, we introduce test-time adaptation (TTA), which performs lightweight unsupervised optimization during inference without access to ground-truth undistorted images, leading to improved performance on OOD data. We validate SACRED under multiple distribution-shift settings and demonstrate that it outperforms existing calibration scan-free fMRI distortion correction methods on both in-distribution (ID) and OOD datasets. The main contributions of this study are summarized as follows:

\begin{itemize}
\item[1)] We propose SACRED, a calibration scan-free fMRI susceptibility-induced distortion correction (SDC) method that requires only routinely acquired T1w images and unidirectional PE BOLD images.
\item[2)] We introduce an image translation-based registration framework that leverages an INN and MIND-SSC to bridge the contrast gap between BOLD and T1w images, enabling the use of standard mono-contrast similarity metrics while preserving structural consistency.
\item[3)] SACRED demonstrates improved robustness under various distribution shifts, enabled by TTA.
\item[4)] To the best of our knowledge, this study presents the first comprehensive comparative evaluation of nearly all representative publicly available calibration scan-free fMRI distortion correction methods.
\end{itemize}

\section{Related Work}

\subsection{Calibration Scan-Free fMRI Susceptibility Distortion Correction}

Calibration scan-free fMRI SDC has been studied to correct distortions in BOLD images when calibration scans, such as a B0 field map or reverse PE images, are not available~\cite{studholme_accurate_2000, gholipour_cross-validation_2008, kybic2000unwarping, esteban_fmriprep_2019, montez_using_2023, yu2023distortion, jimeno2024gdcnet}. Early calibration scan-free approaches addressed the contrast mismatch between BOLD and anatomical images through either image preprocessing or multimodal similarity metrics. For example, high-pass filtering and histogram equalization were applied to both images to reduce contrast differences and improve their similarity~\cite{kybic2000unwarping}, whereas mutual information-based nonlinear registration was investigated to align BOLD images to anatomical images~\cite{studholme_accurate_2000, gholipour_cross-validation_2008}. Since anatomical images exhibit minimal geometric distortion, they can serve as references for aligning BOLD images and correcting distortions. However, several limitations of mutual information have been reported for multi-contrast registration~\cite{pluim2000image}, and because mutual information is inherently a global measure, it can lead to suboptimal results in local estimation~\cite{heinrich2012mind}.

Subsequent calibration scan-free methods continued to reduce the contrast gap between anatomical and BOLD images using more explicit anatomical-to-BOLD contrast transformations~\cite{esteban_fmriprep_2019, montez_using_2023}. In fMRIPrep~\cite{esteban_fmriprep_2019}, anatomical T1w images were intensity-inverted to heuristically approximate BOLD contrast, thereby facilitating registration. However, such simple preprocessing strategies are limited in their ability to model the complex relationships between anatomical and BOLD image contrasts. More recently, Montez et al.~\cite{montez_using_2023} synthesized an anatomical reference with BOLD-like contrast by decomposing T1w and T2-weighted (T2w) images into radial basis functions and nonlinearly mapping them to BOLD images. However, this contrast mapping approach relies on the availability of both T1w and T2w images to provide sufficiently diverse basis functions, whereas T2w images are not always guaranteed in routine protocols.

Recent advances in deep learning have motivated neural-network-based approaches for distortion correction~\cite{yu2023distortion, jimeno2024gdcnet}. Yu et al.~\cite{yu2023distortion} trained a U-Net~\cite{ronneberger2015u} in a supervised manner to synthesize undistorted BOLD images from distorted BOLD and T1w images, using TOPUP-corrected BOLD images as ground truth. The synthesized images were subsequently used, together with the distorted BOLD images, as inputs to TOPUP for VDM estimation. However, supervised learning approaches are sensitive to distribution shifts arising from variations in scanner hardware, acquisition protocols, or subject populations~\cite{su2024navigating}. As a result, in OOD datasets, synthesized images from pretrained U-Nets may exhibit contrast inconsistencies relative to the distorted BOLD images, potentially leading to inaccurate VDM estimation. Alternatively, Jimeno et al.~\cite{jimeno2024gdcnet} employed VoxelMorph~\cite{balakrishnan2019voxelmorph}, an unsupervised deformable registration network, to register distorted BOLD images directly to T1w images while constraining deformation to the PE direction. Nevertheless, the substantial contrast mismatch between these images remains a challenge because commonly used similarity metrics, such as mean squared error (MSE) or local normalized cross-correlation (LNCC), are ineffective for images with disparate intensity distributions.

\subsection{Multi-Contrast Medical Image Registration}
Multi-contrast medical image registration has been widely studied to mitigate optimization challenges arising from contrast mismatch between images. Early approaches relied on maximizing mutual information~\cite{pluim2000image, maes2002multimodality, pluim2003mutual}, which demonstrated strong performance for rigid alignment but often struggled to accurately estimate local deformations. Subsequently, self-similarity-based methods were introduced~\cite{heinrich2012mind, heinrich2013towards}, enabling the capture of local structural patterns that are less sensitive to image contrast. More recently, several studies have addressed multi-contrast registration by transforming it into a mono-contrast problem using image-to-image translation networks~\cite{qin2019unsupervised, chen2022unsupervised, guo2024unsupervised}. In particular, Guo et al.~\cite{guo2024unsupervised} demonstrated the potential of INNs~\cite{dinh2014nice, dinh_density_2017, kingma_glow_2018} as an image-translation backbone, facilitating improved spatial consistency by avoiding information loss through bijective mappings.

\begin{figure*}[!ht]
    \centering
    \includegraphics[width=0.8\textwidth]{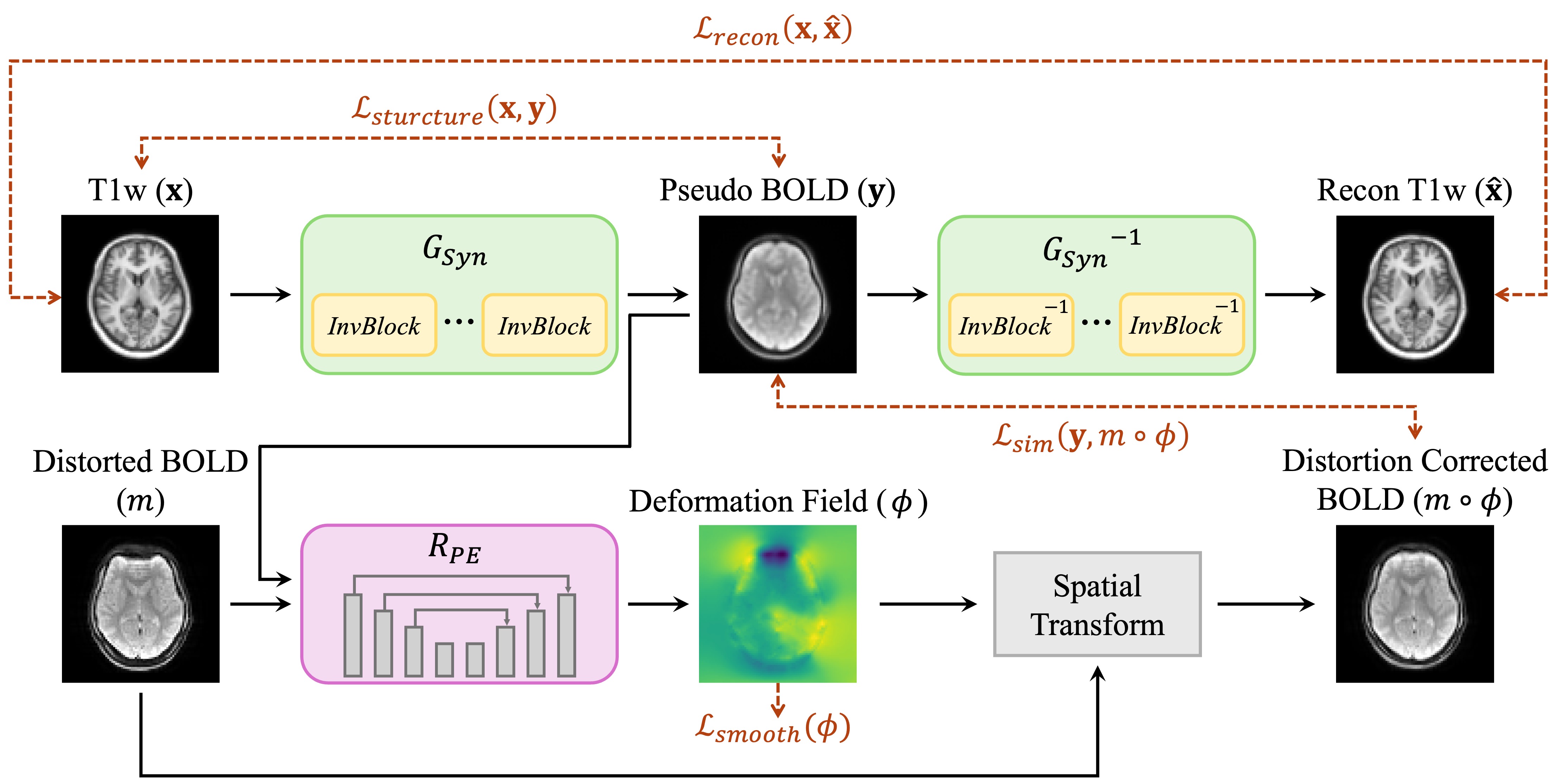}
    \caption{Overview of the proposed susceptibility distortion correction framework.}
    \label{fig:overview}
\end{figure*}

\section{Methods}

\subsection{Framework Overview}
An overview of the proposed SACRED framework is illustrated in Fig.~\ref{fig:overview}. The framework consists of two neural networks: an image translation network ($G_{Syn}$ in Fig.~\ref{fig:overview}) and an image registration network ($R_{PE}$ in Fig.~\ref{fig:overview}). By jointly training these two networks, the registration network can be optimized using a mono-contrast similarity objective, while the image translation network is trained without requiring ground-truth undistorted BOLD images.

Specifically, given a T1w image, the forward transformation of the image translation network $G_{Syn}$ synthesizes a pseudo BOLD image, whereas the reverse transformation reconstructs the T1w image from the synthesized pseudo BOLD image. The image registration network $R_{PE}$ takes the distorted BOLD image and the pseudo BOLD image as inputs and estimates a deformation field $\phi$ to correct geometric distortions. Using the estimated deformation field, the distorted BOLD image is spatially transformed to align with the pseudo BOLD image, with the deformation constrained to the PE direction.

\subsection{Image Translation Network}
\begin{figure}[!ht]
    \centering
    \includegraphics[width=\linewidth]{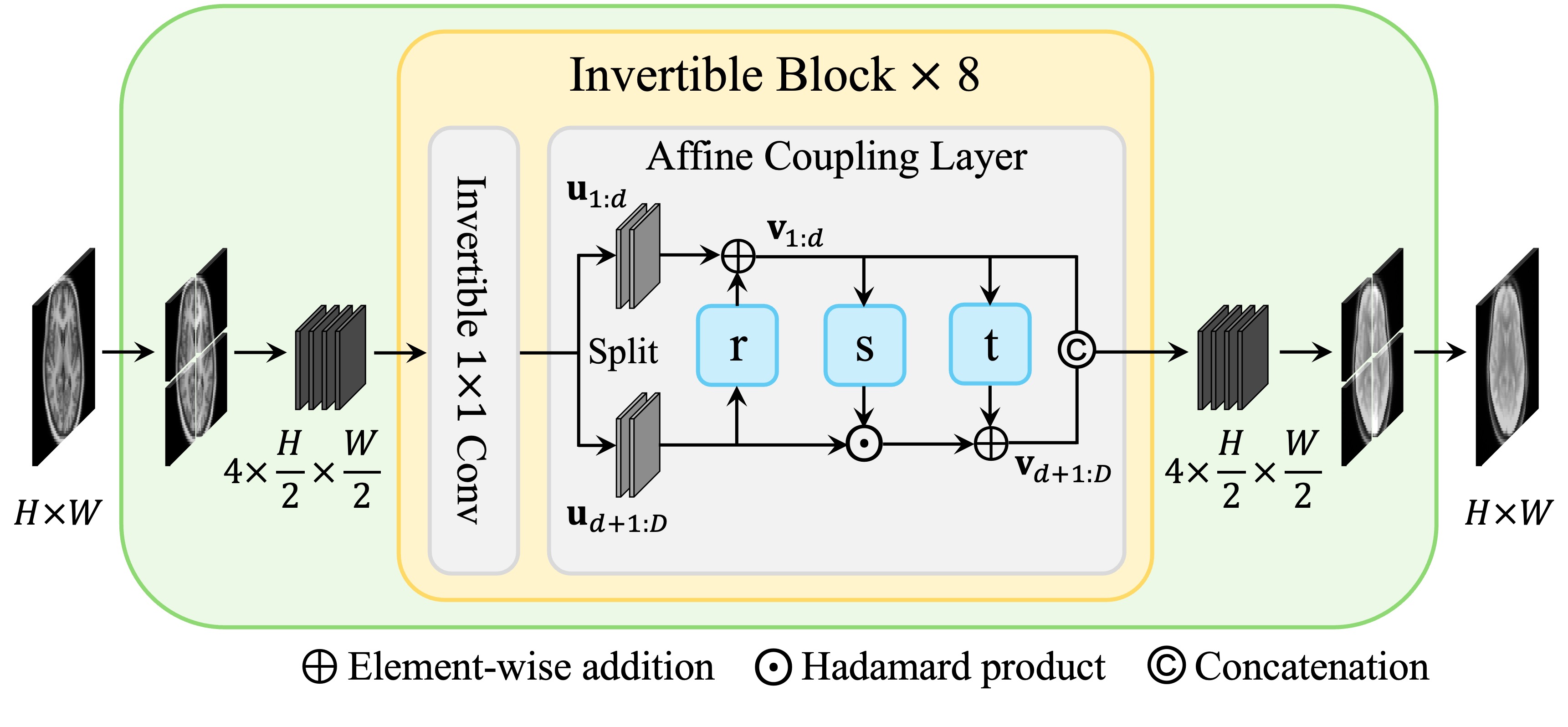}
    \caption{Architecture of the image translation network ($G_{Syn}$). The input T1-weighted image is split into four non-overlapping spatial patches, which are stacked along the channel dimension. The resulting representation is then processed by eight invertible blocks for image translation.}
    \label{fig:INN}
\end{figure}

Preserving the anatomical structure of the input T1w image during image translation is important for synthesizing an undistorted pseudo BOLD image. To this end, we adopted an INN as the backbone of the image translation network, enabling bijective mappings and avoiding information loss during translation. Specifically, to directly map a T1w image to a pseudo BOLD image without introducing a latent distribution, we followed the implementation proposed in~\cite{xing_invertible_2021}. The overall architecture of the image translation network is illustrated in Fig.~\ref{fig:INN}.

Our invertible image translation network is formulated as a mapping $f: \mathcal{X} \rightarrow \mathcal{Y}$, where $\mathcal{X}$ denotes the T1w image domain and $\mathcal{Y}$ denotes the undistorted BOLD image domain. The mapping is implemented using eight invertible blocks $\{f_i\}_{i=0}^{7}$. Given an input T1w image $\mathbf{x}$, the corresponding pseudo BOLD image $\mathbf{y}$ is obtained as
\begin{align}
    \mathbf{y}  = f_7 \circ f_6 \circ \cdots \circ f_1 \circ f_0(\mathbf{x}),
\label{eq:forward}
\end{align}
and the reconstructed T1w image $\mathbf{\hat{x}}$ can be recovered by applying the inverse transformations,
\begin{align}
    \mathbf{\hat{x}} = f_0^{-1} \circ f_1^{-1} \circ \cdots \circ f_6^{-1} \circ f_7^{-1}(\mathbf{y}),
\label{eq:reverse}
\end{align}
where $f_i^{-1}$ denotes the inverse operation of the $i$-th invertible block.

As shown in Fig.~\ref{fig:INN}, each invertible block $f_i$ is composed of an affine coupling layer~\cite{dinh_density_2017}. Given a $D$-dimensional input vector $\mathbf{u}$ and a split index $d$ ($d < D$), the affine coupling layer partitions $\mathbf{u}$ into two sub-vectors. The corresponding output vector $\mathbf{v}$ is computed as
\begin{align}
    \mathbf{v}_{1:d} &= \mathbf{u}_{1:d} + r\!\left(\mathbf{u}_{d+1:D}\right), \\
    \mathbf{v}_{d+1:D} &= \mathbf{u}_{d+1:D} \odot \exp\!\left(s\!\left(\mathbf{v}_{1:d}\right)\right) + t\!\left(\mathbf{v}_{1:d}\right),
\end{align}
where $\odot$ denotes the Hadamard product. The functions $r$, $s$, and $t$ are arbitrary learnable functions implemented using a 5-layer densely connected convolutional block~\cite{huang2017densely} and are not required to be invertible. The inverse mapping can be computed analytically as
\begin{align}
    \mathbf{u}_{d+1:D} &= \left(\mathbf{v}_{d+1:D} - t\!\left(\mathbf{v}_{1:d}\right)\right) \odot \exp\!\left(-s\!\left(\mathbf{v}_{1:d}\right)\right), \\
    \mathbf{u}_{1:d} &= \mathbf{v}_{1:d} - r\!\left(\mathbf{u}_{d+1:D}\right).
\end{align}

In our implementation, a 4-dimensional input vector $\mathbf{u}$ was constructed by partitioning the T1w image $\mathbf{x}$ into four non-overlapping spatial patches and stacking them along the channel dimension, with the split index set to $d=2$ (see Fig.~\ref{fig:INN}). In addition, we employed an invertible $1\times1$ convolution~\cite{kingma_glow_2018} as a permutation layer to reverse the channel order before subsequent affine coupling layers.

\subsection{Image Registration Network}
SDC can be formulated as the estimation of a deformation field that specifies voxel-wise displacements along the PE direction, followed by unwarping the distorted BOLD image using the estimated field. To this end, we employed a U-Net~\cite{Dalca_2018} as the registration backbone to predict a deformation field that aligns the distorted BOLD image with the synthesized pseudo BOLD image.

\begin{figure}[!ht]
    \centering
    \includegraphics[width=\linewidth]{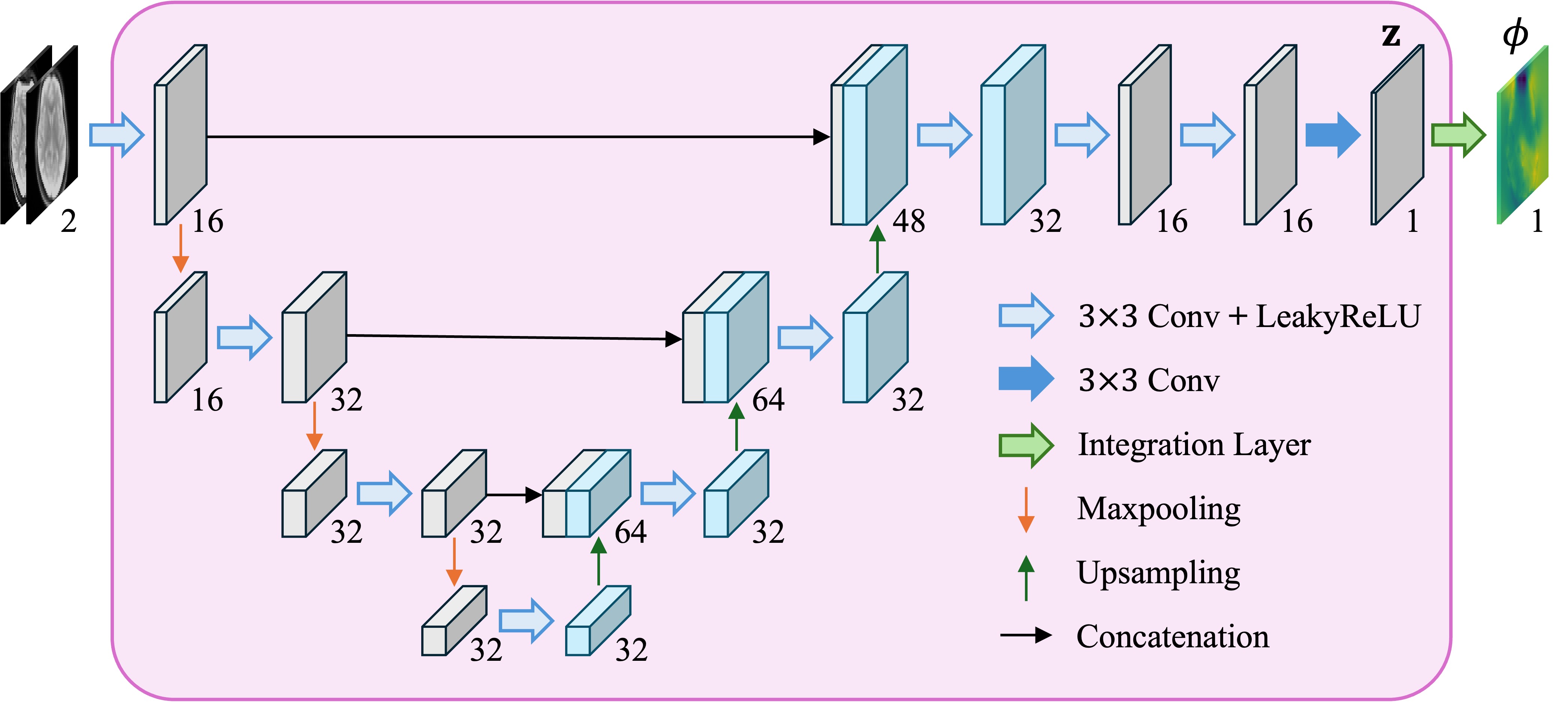}
    \caption{Architecture of the image registration network ($R_{PE}$). Given the distorted BOLD and pseudo BOLD images, the final convolutional layer outputs a velocity field, which is converted into a deformation field via an integration layer.}
    \label{fig:Reg}
\end{figure}

The architecture of the registration network is illustrated in Fig.~\ref{fig:Reg}. The network follows an encoder--decoder design with skip connections. The moving distorted BOLD image is concatenated with the pseudo BOLD image along the channel dimension and provided as input to the network. All convolutional layers are two-dimensional, with a kernel size of $3$ and a stride of $1$.

In the encoding stage, each convolutional block consists of a convolutional layer, a LeakyReLU activation with a negative slope of $0.2$, and a maxpooling layer with kernel size and stride of $2$. This block is repeated three times, progressively reducing the spatial resolution by a factor of eight while increasing the number of feature channels up to $32$. The bottleneck comprises a single convolutional layer followed by a LeakyReLU activation. In the decoding stage, the network alternates between upsampling, feature concatenation from the encoder via skip connections, and convolutional operations to gradually restore the input spatial resolution.

To ensure topology preservation and stable training, we constrained the predicted transformation to be diffeomorphic. This is achieved using a scaling-and-squaring integration layer~\cite{Dalca_2018}. Given a single-channel velocity field $\mathbf{z}$ along the PE direction predicted by the network and the number of integration steps $K$, the deformation field $\phi$ is computed recursively as
\begin{align}
\phi^{(0)} &= \mathbf{p} + \frac{\mathbf{z}}{2^K}, \\
\phi^{(k+1)} &= \phi^{(k)} \circ \phi^{(k)}, \\
\phi &\triangleq \phi^{(K)},
\end{align}
where $\mathbf{p}$ denotes the identity grid and $\circ$ represents the spatial composition (sampling) operator. In this study, we set $K = 7$, corresponding to seven integration steps.

\subsection{Loss Functions}
As shown in Fig.~\ref{fig:overview}, both the image translation network and the image registration network were jointly optimized with a loss function that consists of four components: structural consistency loss $\mathcal{L}_{\text{structure}}$, image reconstruction loss $\mathcal{L}_{\text{recon}}$, similarity loss $\mathcal{L}_{\text{sim}}$, and smoothing loss $\mathcal{L}_{\text{smooth}}$. The detailed definitions of these losses are provided below.

\subsubsection{Structural Consistency Loss}
To ensure the synthesized pseudo BOLD images maintain the anatomical fidelity of the T1w images, we employed a structural consistency loss, $\mathcal{L}_{\text{structure}}$, based on the MIND-SSC descriptor~\cite{heinrich2013towards}.

In our implementation, we defined a 4-neighborhood configuration $\mathcal{N}$ consisting of top, bottom, left, and right neighbors. To capture local geometry, we calculated the intensity differences between pairs of these neighbors. We specifically selected a set of pairs $\mathcal{P}$ from $\mathcal{N}$ that satisfy a squared Euclidean distance of 2 (i.e., pairs of orthogonal neighbors, such as top and left).

For a selected pair $p = (n_i, n_j) \in \mathcal{P}$ at a spatial location $x$ in an input image $I$, the patch-based squared difference $\mathcal{D}_p$ is first calculated as:
\begin{equation}
    \mathcal{D}_p(I, x) = \frac{1}{|\mathcal{R}|} \sum_{\mathbf{r} \in \mathcal{R}} \left( I(x + \mathbf{r} + n_i) - I(x + \mathbf{r} + n_j) \right)^2,
\end{equation}
where $\mathcal{R}$ denotes a local patch centered at $x$ with radius $r$.

To ensure robustness against local contrast variations, a local variance term $\sigma^2(x)$ is estimated as the mean of the minimum-subtracted patch differences across all pairs:
\begin{equation}
    \sigma^2(x) = \frac{1}{|\mathcal{P}|} \sum_{p \in \mathcal{P}} \left( \mathcal{D}_p(I, x) - \min_{q \in \mathcal{P}} \mathcal{D}_q(I, x) \right).
\end{equation}

Based on these terms, the MIND-SSC descriptor element $\mathcal{S}$ for the pair $p$ is computed as:
\begin{align}
    \mathcal{S}(I, x, p) = \exp \left( - \frac{\mathcal{D}_p(I, x) - \min_{q \in \mathcal{P}} \mathcal{D}_q(I, x)}{\sigma^2(x)} \right).
    \label{eq:mind_ssc}
\end{align}

Finally, the structural consistency loss is defined as the MSE over all pixels and all neighbor pairs:
\begin{align}
    \mathcal{L}_{\text{structure}} = \frac{1}{N \cdot |\mathcal{P}|} \sum_{x} \sum_{p} \left( \mathcal{S}(\mathbf{y}, x, p) - \mathcal{S}(\mathbf{x}, x, p) \right)^2,
\end{align}
where $N$ denotes the total number of pixels in the image, and $\mathbf{x}$ and $\mathbf{y}$ are the T1w image and the corresponding pseudo BOLD image obtained by \eqref{eq:forward}, respectively.

\subsubsection{Image Reconstruction Loss}
To leverage the invertibility of the image translation network, we incorporate an image reconstruction loss. This loss enforces that the input image is recoverable from its translated representation. We compute the $L_1$ distance between the original input T1w image $\mathbf{x}$ and the reconstructed image $\mathbf{\hat{x}}$, which is obtained via the inverse mapping defined in \eqref{eq:reverse}:
\begin{align}
    \mathcal{L}_{\text{recon}} = ||\mathbf{x} - \mathbf{\hat{x}}||_1.
\end{align}

\subsubsection{Similarity Loss}
The synthesized pseudo BOLD image enables the use of a mono-contrast similarity metric because it exhibits contrast characteristics similar to those of the BOLD image. We utilize the $L_1$ loss to minimize pixel-wise intensity differences between the pseudo BOLD image and the distortion-corrected BOLD image:
\begin{align}
    \mathcal{L}_{\text{sim}} = ||\mathbf{y} - m \circ \phi||_1,
\end{align}
where $\mathbf{y}$ denotes the synthesized pseudo BOLD image and $m$ represents the input distorted BOLD image. $\phi$ is the predicted deformation field, and $\circ$ denotes the spatial resampling operation. Consequently, the term $m \circ \phi$ corresponds to the distortion-corrected BOLD image aligned with $\mathbf{y}$.

\subsubsection{Smoothing Loss}
To prevent folding and ensure spatial smoothness in the deformation field, we apply a regularization term. This loss penalizes sudden changes in the deformation field by minimizing the $L_2$ norm of its spatial gradients $\nabla \phi$:
\begin{align}
    \mathcal{L}_{\text{smooth}} = ||\nabla\phi||^2_2.
\end{align}

\subsubsection{Total Loss}
The final objective function for joint optimization is a weighted sum of the four loss components discussed above:
\begin{align}
    \mathcal{L}_{\text{total}} = \lambda_1 \cdot \mathcal{L}_{\text{structure}} + \lambda_2 \cdot \mathcal{L}_{\text{recon}} + \lambda_3 \cdot \mathcal{L}_{\text{sim}} + \lambda_4 \cdot \mathcal{L}_{\text{smooth}},
\end{align}
where $\lambda_1, \lambda_2, \lambda_3,$ and $\lambda_4$ are hyperparameters that balance the contribution of each term. In our experiments, we empirically set these weights to $\lambda_1=0.5$, $\lambda_2=0.1$, $\lambda_3=1$, and $\lambda_4=5$.

\begin{table*}[!ht]
\centering
\caption{Summary of the selected subjects and acquisition parameters for the datasets used in this study.}
\label{tab:datasets}
\renewcommand{\arraystretch}{1.2}
\resizebox{\textwidth}{!}{%
\begin{tabular}{lcccccccc}
\toprule[1.2pt]
\multirow{2}{*}{\textbf{Dataset}} &
\multirow{2}{*}{\textbf{OpenNeuro ID}} &
\multirow{2}{*}{\textbf{Vendor}} &
\multirow{2}{*}{\textbf{Field Strength (T)}} &
\multirow{2}{*}{\textbf{Subjects}} &
\multirow{2}{*}{\textbf{Runs}} &
\multicolumn{2}{c}{\textbf{Resolution (mm)}} &
\multirow{2}{*}{\textbf{Age (y)}} \\
\cline{7-8}
& & & & & &
\textbf{T1w} & \textbf{BOLD} & \\
\midrule
NIMH-RV & ds005752 & GE & 3 & 138 & 2 &
$1.0\times1.0\times1.0$ &
$3.0\times3.0\times3.0$ &
18--89 \\

SUDMEX-TMS & ds003037 & Philips & 3 & 53 & 2 &
$1.0\times1.0\times1.0$ &
$3.0\times3.0\times3.3$ &
18--48 \\

QTAB & ds004146 & Siemens & 3 & 50 & 2 &
$0.8\times0.8\times0.8$ &
$2.0\times2.0\times2.0$ &
9--14 \\
\bottomrule[1.2pt]
\end{tabular}
}
\end{table*}

\subsection{Test-Time Adaptation}
\begin{figure}[!ht]
    \centering
    \includegraphics[width=\linewidth]{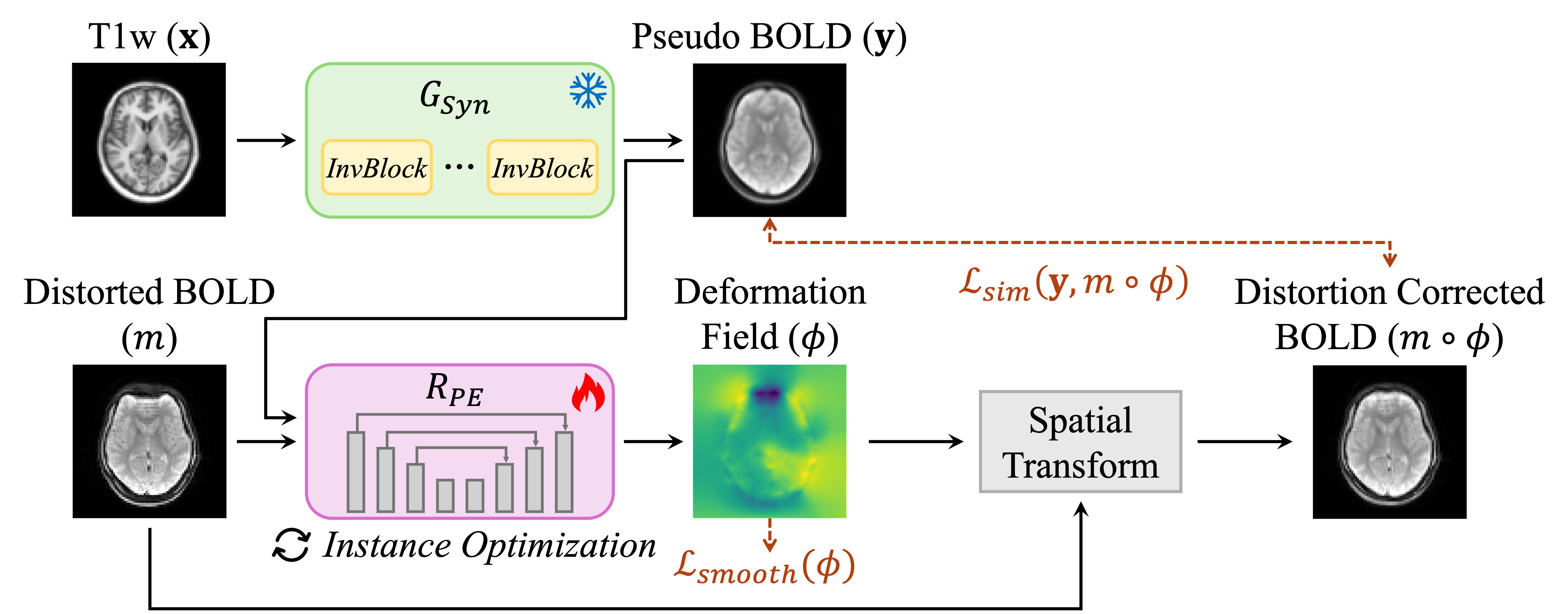}
    \caption{Overview of the proposed test-time adaptation. At this stage, only the registration network ($R_{PE}$) is optimized in an instance-specific manner.}
    \label{fig:TTA}
\end{figure}

Learning-based methods are known to be sensitive to distribution shifts, often resulting in suboptimal performance on unseen datasets~\cite{su2024navigating}. To mitigate this issue, we propose a simple TTA strategy that can be efficiently performed for each test instance. The details of our TTA strategy are illustrated in Fig.~\ref{fig:TTA}.

Given a test T1w image $\mathbf{x}$, the pre-trained image translation network first generates a pseudo BOLD image $\mathbf{y}$. Then, using the input distorted BOLD image $m$ and the generated pseudo BOLD image $\mathbf{y}$, we iteratively optimize only the pre-trained registration network in an instance-specific manner, while keeping the image translation network frozen. The adaptation objective consists of a similarity loss $\mathcal{L}_{\text{sim}}$ and a smoothing loss $\mathcal{L}_{\text{smooth}}$.

During this adaptation phase, we adopt LNCC for $\mathcal{L}_{\text{sim}}$ instead of the $L_{1}$ loss used during training. This design reflects the different roles of the similarity loss in the initial joint training and adaptation stages. During joint training, the image translation network must be explicitly supervised by a pixel-wise loss to generate a faithful pseudo BOLD image. In contrast, during adaptation, the image translation network is fixed, and the objective is solely to refine the deformation field by updating the registration network. Therefore, LNCC can be used as a robust similarity measure for registration refinement without affecting image synthesis.

More specifically, to account for both positive and negative correlations, we employ squared LNCC~\cite{demir2024multigradicon}. This metric improves robustness to contrast discrepancies between the synthesized pseudo BOLD image and the real distorted BOLD image, which may occur when the translation network encounters distribution shifts in the input data. The adaptation loss is defined as:
\begin{align}
    \mathcal{L}_{\text{adap}} = 1 - \text{LNCC}^2(\mathbf{y}, m\circ\phi) + \lambda \cdot ||\nabla\phi||^2_2,
\end{align}
where $\lambda$ is empirically set to 10.

\section{Experimental Setup}
\subsection{Datasets}

Three publicly available datasets~\cite{nugent2022nimh, angeles2024mexican, strike2023queensland} from OpenNeuro~\cite{markiewicz2021openneuro} were used in this study, as summarized in Table~\ref{tab:datasets}. These datasets were selected because each includes two BOLD runs acquired with forward and reverse PE. The NIMH-RV dataset~\cite{nugent2022nimh} served as the ID dataset for training and validation, whereas the SUDMEX-TMS~\cite{angeles2024mexican} and QTAB~\cite{strike2023queensland} datasets were used for independent OOD evaluation. Specifically, SUDMEX-TMS was used to assess robustness to scanner shifts, whereas QTAB was used to evaluate robustness to both scanner and population (age) shifts. The ID dataset was acquired on a GE scanner, while the OOD datasets were acquired on different platforms, with SUDMEX-TMS collected on a Philips scanner and QTAB on a Siemens scanner.

Dataset-specific inclusion criteria were applied based on the availability of required image contrasts. For the NIMH-RV dataset, subjects possessing a complete set of T1w, T2w, and forward and reverse PE BOLD images were included, as T2w images were required for evaluating one baseline method. Consequently, all eligible subjects from the NIMH-RV dataset were retained. In contrast, for the SUDMEX-TMS dataset, no subject-level filtering was applied; however, because T2w images were not available for the entire dataset, the baseline method requiring T2w images was excluded from the evaluation. For the QTAB dataset, all subjects possessed reverse PE BOLD and T2w images, and no exclusion based on image availability was necessary. From the 421 available participants, 50 subjects were randomly sampled for evaluation.

In all experiments, forward and reverse PE BOLD images were treated as independent samples. Accordingly, 278 volumes from the NIMH-RV dataset were used for 5-fold cross-validation. For independent OOD testing, 106 volumes from the SUDMEX-TMS dataset and 100 volumes from the sampled QTAB subjects were used. All datasets used in this study were originally collected with informed consent.

\subsection{Preprocessing}
All three datasets underwent an identical preprocessing pipeline. First, motion correction was applied separately to the forward and reverse PE BOLD images using FSL~\cite{JENKINSON2012782}. Susceptibility-induced distortions were then corrected using TOPUP to generate ground-truth undistorted BOLD images. Each of the forward PE, reverse PE, and TOPUP-corrected BOLD image sets was subsequently averaged across time to obtain 3D volumes. For anatomical preprocessing, N4 bias field correction~\cite{5445030} was applied to the T1w images using FreeSurfer~\cite{FISCHL2012774}. The bias-corrected T1w images were then rigidly registered to the corresponding forward and reverse PE BOLD images using boundary-based registration~\cite{greve2009accurate} implemented in FSL.

\subsection{Implementation Details}

The proposed framework was implemented using PyTorch 2.6.0~\cite{paszke2019pytorchimperativestylehighperformance} and MONAI 1.5.0~\cite{cardoso2022monaiopensourceframeworkdeep}. All experiments were conducted on a single NVIDIA RTX A6000 GPU with 48\,GB memory.

To support downsampling operations, input volumes were minimally zero-padded so that all spatial dimensions were divisible by 16, and the padding was removed after inference. Percentile-based intensity normalization was applied by clipping voxel intensities at the 99th percentile and linearly scaling them to the range $[-1, 1]$. As the model processes 3D volumes in a slice-wise manner, the slice dimension was transposed to the batch dimension during training, resulting in an effective batch size of 1.

The image translation and registration networks were jointly trained for 300 epochs using the Adam optimizer with an initial learning rate of $1\times10^{-4}$ and momentum parameters $\beta_1=0.5$ and $\beta_2=0.999$. A linear learning-rate schedule was used, keeping the rate constant for the first half of training and linearly decaying it to zero thereafter.

For TTA, the registration network was optimized in an instance-specific manner for 500 iterations using Adam with a fixed learning rate of $5\times10^{-5}$ and momentum parameters $\beta_1=0.9$ and $\beta_2=0.999$.

\subsection{Baseline Methods}
To evaluate the effectiveness of the proposed SACRED, we compared its performance with four representative calibration scan-free fMRI SDC methods~\cite{esteban_fmriprep_2019, montez_using_2023, yu2023distortion, jimeno2024gdcnet}, covering both conventional and learning-based approaches.

\subsubsection{Conventional Approaches}
Two representative non-learning-based methods were selected. First, SyN-SDC\footnote{\url{https://github.com/nipreps/fmriprep}}, implemented in the widely used fMRI preprocessing pipeline fMRIPrep~\cite{esteban_fmriprep_2019}, was included. SyN-SDC corrects susceptibility-induced distortions by performing symmetric diffeomorphic registration (SyN)~\cite{avants2008symmetric} between the distorted BOLD image and an anatomical reference, while constraining deformation along the PE direction. Second, Synth\footnote{\url{https://gitlab.com/vanandrew/omni}}~\cite{montez_using_2023} was evaluated. Synth also employs SyN-based non-linear registration, in which an undistorted BOLD-like image is synthesized from both T1-weighted and T2-weighted anatomical images and subsequently used as the registration target for distortion correction.

\subsubsection{Learning-Based Approaches}
We further compared our technique with two recent learning-based SDC methods. First, SynBOLD-DisCo\footnote{\url{https://github.com/MASILab/SynBOLD-DisCo}}~\cite{yu2023distortion} was included. This method trains a U-Net~\cite{ronneberger2015u} in a supervised manner to synthesize an undistorted BOLD image using TOPUP-corrected images as ground truth, and subsequently applies TOPUP using the original distorted BOLD image and the synthesized undistorted counterpart. Second, GDCNet\footnote{\url{https://github.com/imr-framework/gdcnet}}~\cite{jimeno2024gdcnet} was evaluated. GDCNet employs a VoxelMorph-based deformable registration framework~\cite{balakrishnan2019voxelmorph} to directly align distorted BOLD images to anatomical T1-weighted images while constraining the deformation to the PE direction. For a fair comparison, all learning-based methods were trained from scratch using the same NIMH-RV dataset as the proposed framework. Each model was trained for 300 epochs to ensure convergence.

\subsection{Evaluation Metrics}
Quantitative evaluation of calibration scan-free SDC was conducted using three widely adopted image similarity metrics. Normalized root mean squared error (NRMSE) and peak signal-to-noise ratio (PSNR) were used to assess voxel-wise intensity fidelity, while the structural similarity index measure (SSIM) was employed to evaluate structural alignment with the reference. For all evaluations, TOPUP-corrected BOLD images served as the reference standard.

\begin{figure*}[!ht]
    \centering
    \includegraphics[width=0.8\textwidth]{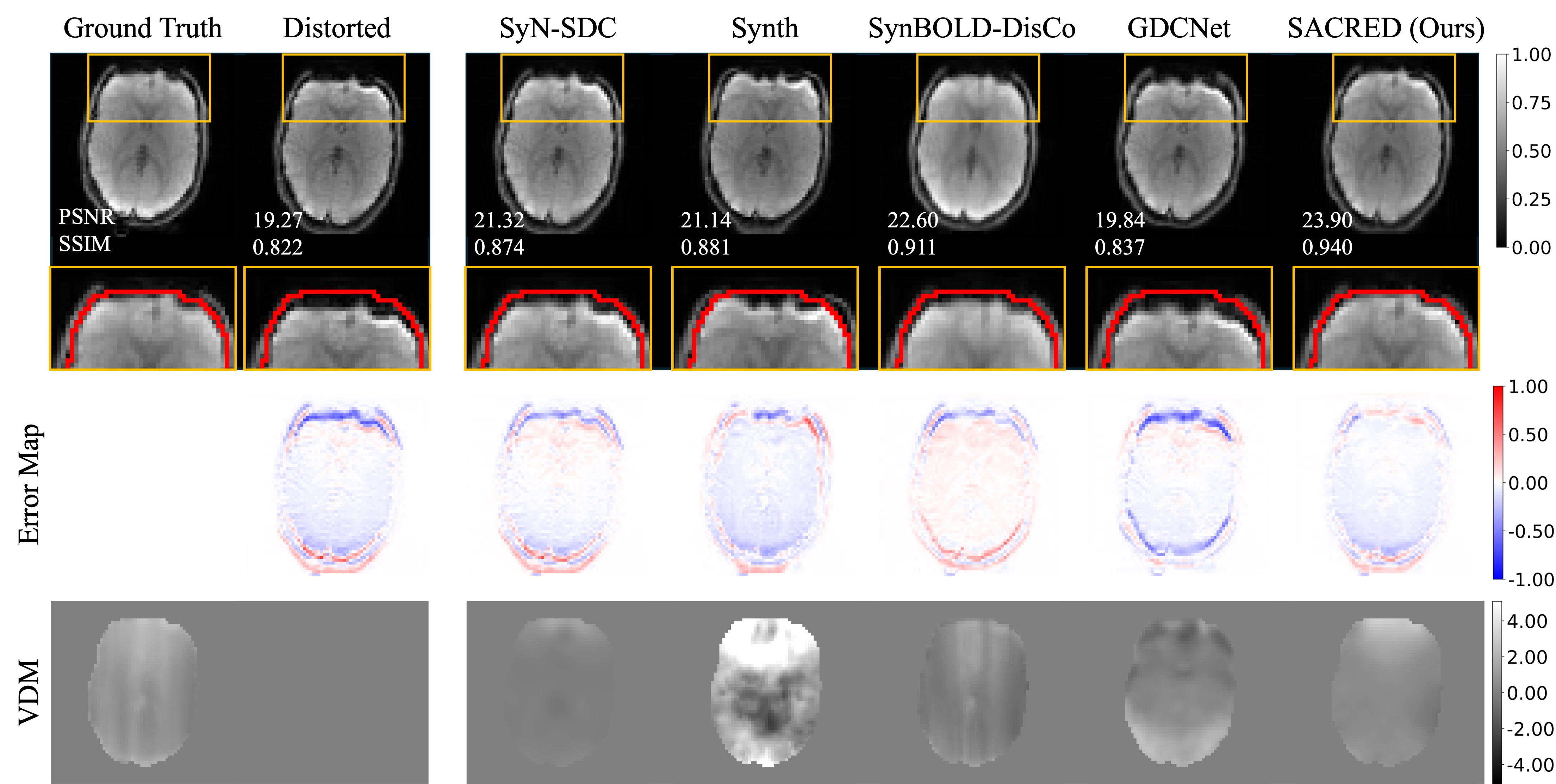}
    \caption{Qualitative results on the ID NIMH-RV dataset. Columns 1--2 show the ground-truth TOPUP-corrected BOLD image and the original distorted BOLD image, respectively. Columns 3--7 present distortion-corrected BOLD images produced by each method. The first row displays the corrected images, with zoomed views of the anterior frontal region highlighted by orange boxes. Red outlines indicate the boundary of the ground-truth brain mask. The second row shows voxel-wise error maps computed with respect to the ground-truth image, where high positive values (+1) indicate stretched regions and negative values (–1) indicate compressed regions. The third row shows voxel displacement maps (VDM) estimated by each method.}
    \label{fig:NIMH-RV}
\end{figure*}

\section{Results}

\subsection{In-Distribution Test Results}

Quantitative results on the NIMH-RV dataset are summarized in Table~\ref{tab:NIMH-RV}. Compared with uncorrected distorted BOLD images, SACRED reduced NRMSE by 22.75\% and increased PSNR and SSIM by 9.49\% and 5.87\%, respectively. SACRED significantly outperformed all competing methods in terms of NRMSE and SSIM. For PSNR, SACRED achieved the highest mean value, with statistically significant improvements over all methods except SynBOLD-DisCo, for which the difference was marginal and not statistically significant. SynBOLD-DisCo achieved the second-best performance, likely due to its supervised training with TOPUP-corrected ground-truth images.

Fig.~\ref{fig:NIMH-RV} shows representative corrected images, voxel-wise error maps computed relative to the ground truth, and VDMs estimated by each method. A slice near the orbitofrontal cortex was selected because this region is prone to severe susceptibility-induced distortions caused by air--tissue interfaces in the sinuses~\cite{ojemann1997anatomic}. Zoomed views of the anterior frontal region are highlighted, with the ground-truth brain mask boundary obtained using HD-BET~\cite{isensee2019automated} overlaid. SACRED exhibited the closest anatomical alignment and the lowest residual errors among all methods. Compared with the baseline methods, the VDM estimated by SACRED showed displacement values and spatial distributions that were largely consistent with the ground-truth VDM, while avoiding the stripe-like patterns observed in the ground-truth VDM. Consequently, SACRED produced a smoother VDM estimate. In contrast, SynBOLD-DisCo, which estimates the VDM using TOPUP, reproduced the stripe-like patterns observed in the ground-truth VDM but showed overall lower displacement values.

\begin{table}[!ht]
\centering
\caption{Quantitative comparison on NIMH-RV dataset. Values are reported as mean $\pm$ standard deviation. $^*$ indicates that SACRED significantly outperformed the corresponding method ($p < 0.01$, Wilcoxon signed-rank test).}
\label{tab:NIMH-RV}
\renewcommand{\arraystretch}{1.2}
\resizebox{\columnwidth}{!}{
\begin{tabular}{lccc}
\toprule[1.2pt]
\textbf{Method} & \textbf{NRMSE} $\downarrow$ & \textbf{PSNR} $\uparrow$ & \textbf{SSIM} $\uparrow$ \\
\midrule
Distorted      & $0.0589^*$ {\color{gray}\scriptsize $\pm 0.0130$} & $24.7915^*$ {\color{gray}\scriptsize $\pm 1.8097$} & $0.8591^*$ {\color{gray}\scriptsize $\pm 0.0538$} \\
SyN-SDC        & $0.0528^*$ {\color{gray}\scriptsize $\pm 0.0127$} & $25.7688^*$ {\color{gray}\scriptsize $\pm 1.9238$} & $0.8744^*$ {\color{gray}\scriptsize $\pm 0.0513$} \\
Synth          & $0.0703^*$ {\color{gray}\scriptsize $\pm 0.0144$} & $23.2542^*$ {\color{gray}\scriptsize $\pm 1.8331$} & $0.7668^*$ {\color{gray}\scriptsize $\pm 0.0635$} \\
SynBOLD-DisCo  & $0.0496^*$ {\color{gray}\scriptsize $\pm 0.0186$} & $26.6267$   {\color{gray}\scriptsize $\pm 3.0024$} & $0.8870^*$ {\color{gray}\scriptsize $\pm 0.0725$} \\
GDCNet         & $0.0815^*$ {\color{gray}\scriptsize $\pm 0.0107$} & $21.8470^*$ {\color{gray}\scriptsize $\pm 1.1339$} & $0.7783^*$ {\color{gray}\scriptsize $\pm 0.0460$} \\
\midrule
\textbf{SACRED (Ours)} & $\mathbf{0.0455}$ {\color{gray}\scriptsize $\pm 0.0131$} & $\mathbf{27.1434}$ {\color{gray}\scriptsize $\pm 2.2227$} & $\mathbf{0.9095}$ {\color{gray}\scriptsize $\pm 0.0519$} \\
\bottomrule[1.2pt]
\end{tabular}
}
\end{table}

\begin{figure*}[!ht]
    \centering
    \includegraphics[width=0.8\textwidth]{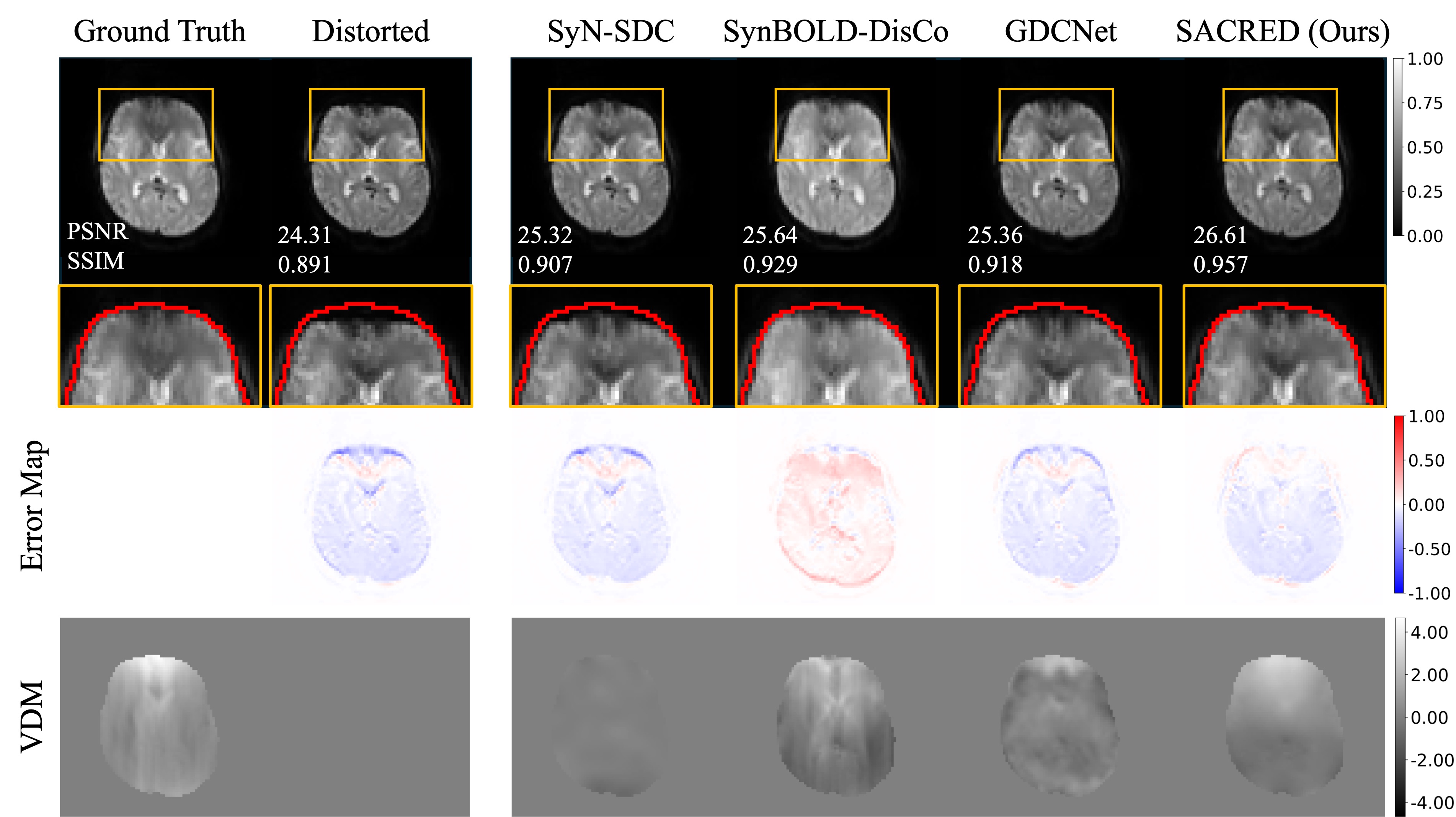}
    \caption{Qualitative results on the OOD SUDMEX-TMS dataset. Columns 1--2 show the ground-truth TOPUP-corrected BOLD image and the original distorted BOLD image, respectively. Columns 3--6 present distortion-corrected BOLD images produced by each method. The first row displays the corrected images, with zoomed views of the anterior frontal region highlighted by orange boxes. Red outlines indicate the boundary of the ground-truth brain mask. The second row shows voxel-wise error maps computed with respect to the ground-truth image, where high positive values (+1) indicate stretched regions and negative values (–1) indicate compressed regions. The third row shows voxel displacement maps (VDM) estimated by each method.}
    \label{fig:SUDMEX-TMS}
\end{figure*}

\subsection{Out-of-Distribution Test Results}

\subsubsection{SUDMEX-TMS Dataset}

Synth was excluded from this evaluation due to the absence of T2w images in the SUDMEX-TMS dataset. Quantitative results are summarized in Table~\ref{tab:SUDMEX-TMS}. As this dataset was acquired on a different scanner platform, it enables evaluation under scanner-induced distribution shifts.

Compared with the uncorrected distorted BOLD images, SACRED reduced NRMSE by 7.0\% and increased PSNR and SSIM by 2.7\% and 2.1\%, respectively. SACRED significantly outperformed all competing methods across all metrics. In contrast to the ID results, SyN-SDC achieved the second-best performance, highlighting the sensitivity of learning-based approaches to scanner-induced distribution shifts. Specifically, SynBOLD-DisCo and GDCNet showed worse quantitative performance than the original distorted BOLD images. Despite this, SACRED maintained stable performance on this OOD dataset, demonstrating improved robustness partly attributable to TTA.

Representative qualitative results, error maps, and estimated VDMs are shown in Fig.~\ref{fig:SUDMEX-TMS}. As in the ID evaluation, a slice near the orbitofrontal cortex was selected. While both SynBOLD-DisCo and SACRED improved alignment compared with the other methods, SynBOLD-DisCo exhibited contrast inconsistencies, potentially due to mismatches between the synthesized and distorted BOLD images jointly used in TOPUP. Regarding the VDMs, the VDM estimated by SACRED showed displacement values and spatial distributions closest to the ground-truth VDM, while avoiding the stripe-like patterns observed in the ground truth and producing a smoother VDM estimate.

\begin{figure*}[!ht]
    \centering
    \includegraphics[width=0.8\textwidth]{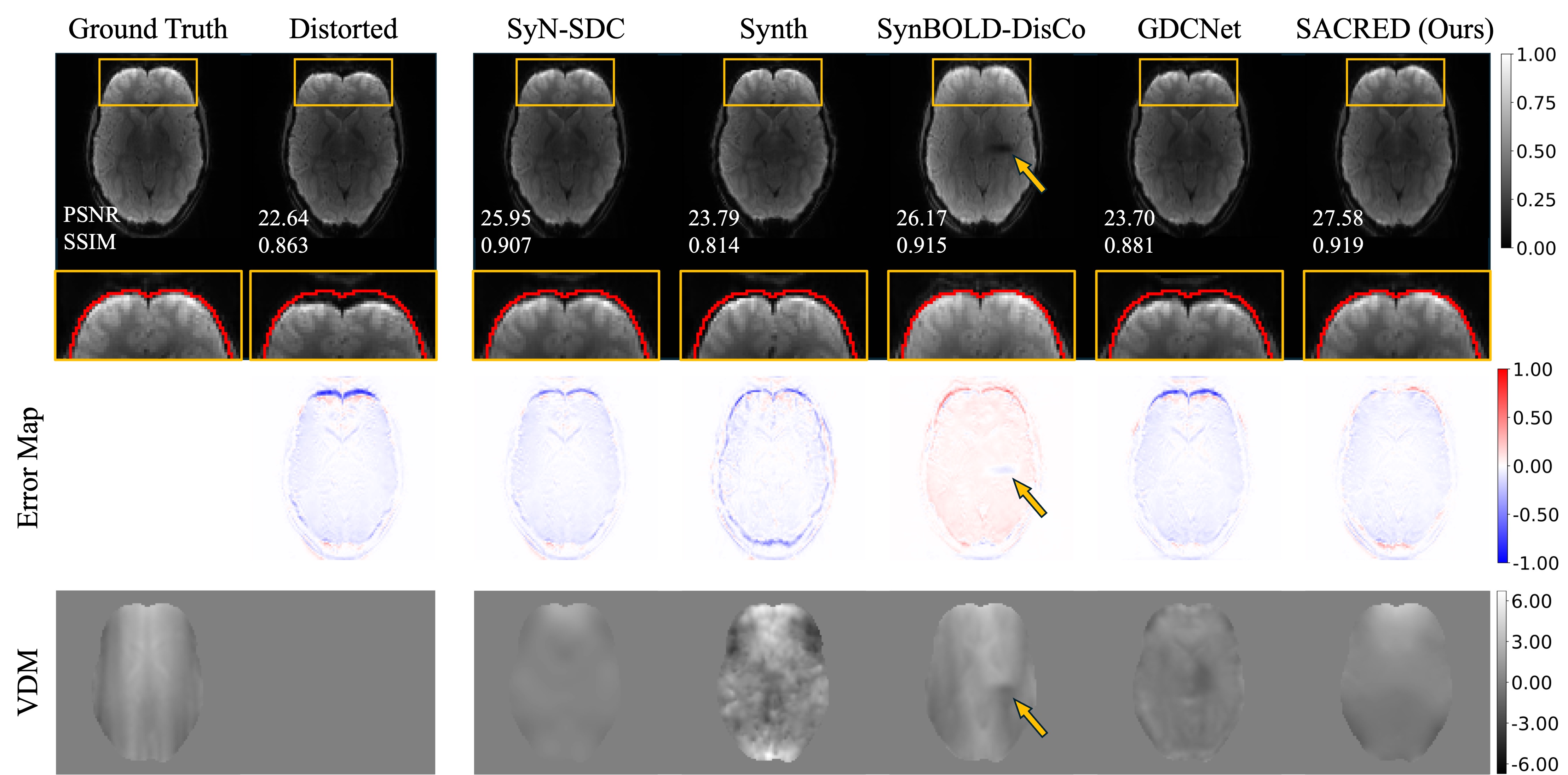}
    \caption{Qualitative results on the OOD QTAB dataset. Columns 1--2 show the ground-truth TOPUP-corrected BOLD image and the original distorted BOLD image, respectively. Columns 3--7 present distortion-corrected BOLD images produced by each method. The first row displays the corrected images, with zoomed views of the anterior frontal region highlighted by orange boxes. Red outlines indicate the boundary of the ground-truth brain mask. An unexpected artifact introduced after correction is highlighted by an orange arrow. The second row shows voxel-wise error maps computed with respect to the ground-truth image, where high positive values (+1) indicate stretched regions and negative values (–1) indicate compressed regions. The third row shows voxel displacement maps (VDM) estimated by each method.}
    \label{fig:QTAB}
\end{figure*}

\begin{figure}[!ht]
    \centering
    \includegraphics[width=\linewidth]{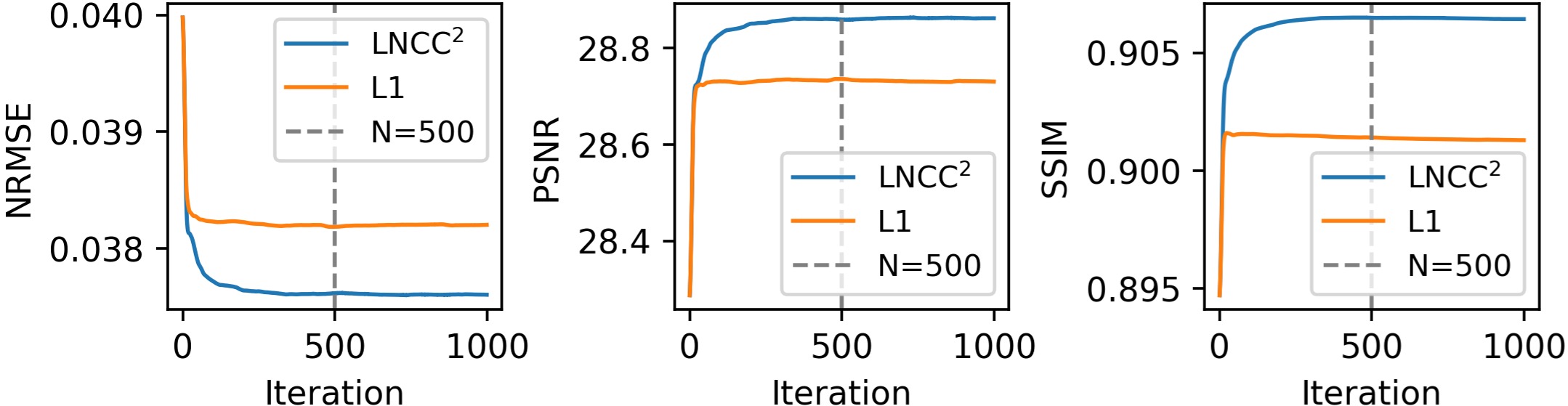}
    \caption{Effect of the similarity loss function on test-time adaptation performance. The blue and orange curves indicate the performance trends obtained using the proposed $\text{LNCC}^2$ loss for TTA and the $L_1$ loss, which was used for initial training and evaluated here as a TTA objective, respectively. The results were computed by applying a moving average over all volumes from the three datasets. The vertical dashed line denotes the selected number of TTA iterations, $N=500$.}
    \label{fig:tta_analysis}
\end{figure}

\begin{table}[!ht]
\centering
\caption{Quantitative comparison on the SUDMEX-TMS dataset. Values are reported as mean $\pm$ standard deviation. $^*$ indicates that SACRED significantly outperformed the corresponding method ($p < 0.01$, Wilcoxon signed-rank test).}
\label{tab:SUDMEX-TMS}
\renewcommand{\arraystretch}{1.2}
\resizebox{\columnwidth}{!}{%
\begin{tabular}{lccc}
\toprule[1.2pt]
\textbf{Method} & \textbf{NRMSE} $\downarrow$ & \textbf{PSNR} $\uparrow$ & \textbf{SSIM} $\uparrow$ \\
\midrule
Distorted      & $0.0502^*$ {\color{gray}\scriptsize $\pm 0.0119$} & $26.2237^*$ {\color{gray}\scriptsize $\pm 2.0085$} & $0.8345^*$ {\color{gray}\scriptsize $\pm 0.0622$} \\
SyN-SDC        & $0.0483^*$ {\color{gray}\scriptsize $\pm 0.0119$} & $26.5693^*$ {\color{gray}\scriptsize $\pm 2.0939$} & $0.8408^*$ {\color{gray}\scriptsize $\pm 0.0616$} \\
SynBOLD-DisCo  & $0.0605^*$ {\color{gray}\scriptsize $\pm 0.0146$} & $24.6078^*$ {\color{gray}\scriptsize $\pm 2.0880$} & $0.8267^*$ {\color{gray}\scriptsize $\pm 0.0679$} \\
GDCNet         & $0.0512^*$ {\color{gray}\scriptsize $\pm 0.0117$} & $26.0300^*$ {\color{gray}\scriptsize $\pm 1.9456$} & $0.8343^*$ {\color{gray}\scriptsize $\pm 0.0653$} \\
\midrule
\textbf{SACRED (Ours)} & $\mathbf{0.0467}$ {\color{gray}\scriptsize $\pm 0.0130$} & $\mathbf{26.9338}$ {\color{gray}\scriptsize $\pm 2.3592$} & $\mathbf{0.8521}$ {\color{gray}\scriptsize $\pm 0.0671$} \\
\bottomrule[1.2pt]
\end{tabular}%
}
\end{table}

\subsubsection{QTAB Dataset}

Quantitative results on the QTAB dataset are summarized in Table~\ref{tab:QTAB}. As an OOD dataset acquired on a different scanner platform and comprising subjects from a distinct age range, QTAB enables evaluation under both scanner and population distribution shifts.

Compared with the uncorrected distorted BOLD images, SACRED reduced NRMSE by approximately 16.43\% and increased PSNR and SSIM by 5.69\% and 2.29\%, respectively. SACRED achieved the lowest NRMSE and highest SSIM among all evaluated methods and significantly outperformed competing approaches on these metrics. For PSNR, SACRED attained the highest mean value, with statistically significant improvements over all methods except SyN-SDC, for which the difference was marginal and not statistically significant.

Representative qualitative results, corresponding voxel-wise error maps, and estimated VDMs are shown in Fig.~\ref{fig:QTAB}. As in the previous evaluations, a slice near the orbitofrontal cortex was selected. Among all methods, SACRED exhibited the closest anatomical alignment with the ground-truth images, as reflected by reduced residual errors in the error maps. In contrast, SynBOLD-DisCo introduced an unexpected artifact after correction, highlighted by the orange arrow in Fig.~\ref{fig:QTAB}, indicating reduced robustness on the OOD dataset. Regarding the VDMs, SynBOLD-DisCo and SACRED showed spatial displacement patterns most similar to that of the ground-truth VDM. However, as highlighted by the orange arrow, SynBOLD-DisCo produced a locally implausible displacement estimate, which spatially corresponded to the correction artifact described above. In contrast, SACRED produced a smoother VDM estimate without such local irregularities.

\begin{table}[!ht]
\centering
\caption{Quantitative comparison on the QTAB dataset. Values are reported as mean $\pm$ standard deviation. $^*$ indicates that SACRED significantly outperformed the corresponding method ($p < 0.01$, Wilcoxon signed-rank test).}
\label{tab:QTAB}
\renewcommand{\arraystretch}{1.2}
\resizebox{\columnwidth}{!}{%
\begin{tabular}{lccc}
\toprule[1.2pt]
\textbf{Method} & \textbf{NRMSE} $\downarrow$ & \textbf{PSNR} $\uparrow$ & \textbf{SSIM} $\uparrow$ \\
\midrule
Distorted      & $0.0414^*$ {\color{gray}\scriptsize $\pm 0.0075$} & $27.7994^*$ {\color{gray}\scriptsize $\pm 1.4840$} & $0.8916^*$ {\color{gray}\scriptsize $\pm 0.0326$} \\
SyN-SDC        & $0.0357^*$ {\color{gray}\scriptsize $\pm 0.0083$} & $29.1514$   {\color{gray}\scriptsize $\pm 1.8645$} & $0.9077^*$ {\color{gray}\scriptsize $\pm 0.0350$} \\
Synth          & $0.0450^*$ {\color{gray}\scriptsize $\pm 0.0077$} & $27.0627^*$ {\color{gray}\scriptsize $\pm 1.4500$} & $0.8543^*$ {\color{gray}\scriptsize $\pm 0.0303$} \\
SynBOLD-DisCo  & $0.0390^*$ {\color{gray}\scriptsize $\pm 0.0098$} & $28.4276^*$ {\color{gray}\scriptsize $\pm 2.0658$} & $0.8901^*$ {\color{gray}\scriptsize $\pm 0.0419$} \\
GDCNet         & $0.0371^*$ {\color{gray}\scriptsize $\pm 0.0075$} & $28.7691^*$ {\color{gray}\scriptsize $\pm 1.5516$} & $0.9015^*$ {\color{gray}\scriptsize $\pm 0.0324$} \\
\midrule
\textbf{SACRED (Ours)} & $\mathbf{0.0346}$ {\color{gray}\scriptsize $\pm 0.0075$} & $\mathbf{29.3818}$ {\color{gray}\scriptsize $\pm 1.6772$} & $\mathbf{0.9120}$ {\color{gray}\scriptsize $\pm 0.0326$} \\
\bottomrule[1.2pt]
\end{tabular}%
}
\end{table}

\begin{table*}[!ht]
\centering
\caption{Ablation studies evaluating the effects of the loss functions and TTA on the NIMH-RV, SUDMEX-TMS, and QTAB datasets. Values are reported as mean $\pm$ standard deviation. $^*$ indicates that the full SACRED model, with all three components enabled, significantly outperformed the corresponding ablated variant ($p < 0.01$, Wilcoxon signed-rank test).}
\label{tab:ablation}
\renewcommand{\arraystretch}{1.2}
\resizebox{\textwidth}{!}{%
\begin{tabular}{cccccccccccc}
\toprule[1.2pt]
\multirow{2}{*}{\large $\mathcal{L}_{\text{recon}}$} & \multirow{2}{*}{\large $\mathcal{L}_{\text{structure}}$} & \multirow{2}{*}{\large TTA} & \multicolumn{3}{c}{\textbf{NIMH-RV}} & \multicolumn{3}{c}{\textbf{SUDMEX-TMS}} & \multicolumn{3}{c}{\textbf{QTAB}} \\
\cmidrule(lr){4-6} \cmidrule(lr){7-9} \cmidrule(lr){10-12}
 & & & \textbf{NRMSE} $\downarrow$ & \textbf{PSNR} $\uparrow$ & \textbf{SSIM} $\uparrow$ & \textbf{NRMSE} $\downarrow$ & \textbf{PSNR} $\uparrow$ & \textbf{SSIM} $\uparrow$ & \textbf{NRMSE} $\downarrow$ & \textbf{PSNR} $\uparrow$ & \textbf{SSIM} $\uparrow$ \\
\midrule
 & & & $0.0546^*$ {\color{gray}\scriptsize $\pm 0.0145$} & $25.5513^*$ {\color{gray}\scriptsize $\pm 2.2396$} & $0.8627^*$ {\color{gray}\scriptsize $\pm 0.0594$} & $0.0591^*$ {\color{gray}\scriptsize $\pm 0.0138$} & $24.8077^*$ {\color{gray}\scriptsize $\pm 2.0448$} & $0.7961^*$ {\color{gray}\scriptsize $\pm 0.0811$} & $0.0510^*$ {\color{gray}\scriptsize $\pm 0.0067$} & $25.9155^*$ {\color{gray}\scriptsize $\pm 1.1175$} & $0.8509^*$ {\color{gray}\scriptsize $\pm 0.0263$} \\
\cmark & & & $0.0545^*$ {\color{gray}\scriptsize $\pm 0.0144$} & $25.5544^*$ {\color{gray}\scriptsize $\pm 2.2255$} & $0.8632^*$ {\color{gray}\scriptsize $\pm 0.0592$} & $0.0583^*$ {\color{gray}\scriptsize $\pm 0.0141$} & $24.9408^*$ {\color{gray}\scriptsize $\pm 2.1250$} & $0.7999^*$ {\color{gray}\scriptsize $\pm 0.0832$} & $0.0508^*$ {\color{gray}\scriptsize $\pm 0.0071$} & $25.9575^*$ {\color{gray}\scriptsize $\pm 1.1853$} & $0.8513^*$ {\color{gray}\scriptsize $\pm 0.0271$} \\
\cmark & \cmark & & $0.0471^*$ {\color{gray}\scriptsize $\pm 0.0130$} & $26.8188^*$ {\color{gray}\scriptsize $\pm 2.1457$} & $0.9008^*$ {\color{gray}\scriptsize $\pm 0.0517$} & $0.0509^*$ {\color{gray}\scriptsize $\pm 0.0134$} & $26.1700^*$ {\color{gray}\scriptsize $\pm 2.2764$} & $0.8326^*$ {\color{gray}\scriptsize $\pm 0.0726$} & $0.0391^*$ {\color{gray}\scriptsize $\pm 0.0066$} & $28.2679^*$ {\color{gray}\scriptsize $\pm 1.3878$} & $0.8881^*$ {\color{gray}\scriptsize $\pm 0.0296$} \\
\cmark & \cmark & \cmark & $\mathbf{0.0455}$ {\color{gray}\scriptsize $\pm 0.0131$} & $\mathbf{27.1434}$ {\color{gray}\scriptsize $\pm 2.2227$} & $\mathbf{0.9095}$ {\color{gray}\scriptsize $\pm 0.0519$} & $\mathbf{0.0467}$ {\color{gray}\scriptsize $\pm 0.0130$} & $\mathbf{26.9338}$ {\color{gray}\scriptsize $\pm 2.3592$} & $\mathbf{0.8521}$ {\color{gray}\scriptsize $\pm 0.0671$} & $\mathbf{0.0346}$ {\color{gray}\scriptsize $\pm 0.0075$} & $\mathbf{29.3818}$ {\color{gray}\scriptsize $\pm 1.6772$} & $\mathbf{0.9120}$ {\color{gray}\scriptsize $\pm 0.0326$} \\
\bottomrule[1.2pt]
\end{tabular}%
}
\end{table*}

\begin{table*}[!ht]
\centering
\caption{Ablation studies evaluating the image translation network architecture on the NIMH-RV, SUDMEX-TMS, and QTAB datasets. Values are reported as mean $\pm$ standard deviation. $^*$ indicates that the proposed INN-based architecture significantly outperformed the corresponding ablated networks ($p < 0.01$, Wilcoxon signed-rank test).}
\label{tab:ablation_inn}
\renewcommand{\arraystretch}{1.2}
\resizebox{\textwidth}{!}{%
\begin{tabular}{cccccccccccc}
\toprule[1.2pt]
\multirow{2}{*}{\textbf{Architecture}} & \multirow{2}{*}{\textbf{TTA}} & \multicolumn{3}{c}{\textbf{NIMH-RV}} & \multicolumn{3}{c}{\textbf{SUDMEX-TMS}} & \multicolumn{3}{c}{\textbf{QTAB}} \\
\cmidrule(lr){3-5} \cmidrule(lr){6-8} \cmidrule(lr){9-11}
 & & \textbf{NRMSE} $\downarrow$ & \textbf{PSNR} $\uparrow$ & \textbf{SSIM} $\uparrow$ & \textbf{NRMSE} $\downarrow$ & \textbf{PSNR} $\uparrow$ & \textbf{SSIM} $\uparrow$ & \textbf{NRMSE} $\downarrow$ & \textbf{PSNR} $\uparrow$ & \textbf{SSIM} $\uparrow$ \\
\midrule
\multirow{2}{*}{ResNet Generator}
 & \xmark & $0.0485^*$ {\color{gray}\scriptsize $\pm 0.0132$} & $26.5659^*$ {\color{gray}\scriptsize $\pm 2.1641$} & $0.8962^*$ {\color{gray}\scriptsize $\pm 0.0516$} & $0.0528^*$ {\color{gray}\scriptsize $\pm 0.0140$} & $25.8576^*$ {\color{gray}\scriptsize $\pm 2.2965$} & $0.8204^*$ {\color{gray}\scriptsize $\pm 0.0771$} & $0.0414^*$ {\color{gray}\scriptsize $\pm 0.0066$} & $27.7572^*$ {\color{gray}\scriptsize $\pm 1.2971$} & $0.8817^*$ {\color{gray}\scriptsize $\pm 0.0319$} \\
 & \cmark & $0.0472^*$ {\color{gray}\scriptsize $\pm 0.0132$} & $26.8211^*$ {\color{gray}\scriptsize $\pm 2.2326$} & $0.9041^*$ {\color{gray}\scriptsize $\pm 0.0519$} & $0.0487^*$ {\color{gray}\scriptsize $\pm 0.0124$} & $26.5267^*$ {\color{gray}\scriptsize $\pm 2.1805$} & $0.8429^*$ {\color{gray}\scriptsize $\pm 0.0661$} & $0.0393^*$ {\color{gray}\scriptsize $\pm 0.0070$} & $28.2316^*$ {\color{gray}\scriptsize $\pm 1.4582$} & $0.8983^*$ {\color{gray}\scriptsize $\pm 0.0310$} \\
\midrule
\multirow{2}{*}{U-Net}
 & \xmark & $0.0475^*$ {\color{gray}\scriptsize $\pm 0.0128$} & $26.7440^*$ {\color{gray}\scriptsize $\pm 2.1146$} & $0.8997^*$ {\color{gray}\scriptsize $\pm 0.0508$} & $0.0534^*$ {\color{gray}\scriptsize $\pm 0.0116$} & $25.6554^*$ {\color{gray}\scriptsize $\pm 1.8560$} & $0.8169^*$ {\color{gray}\scriptsize $\pm 0.0636$} & $0.0401^*$ {\color{gray}\scriptsize $\pm 0.0066$} & $28.0463^*$ {\color{gray}\scriptsize $\pm 1.3506$} & $0.8854^*$ {\color{gray}\scriptsize $\pm 0.0320$} \\
 & \cmark & $0.0463^*$ {\color{gray}\scriptsize $\pm 0.0128$} & $26.9791^*$ {\color{gray}\scriptsize $\pm 2.1441$} & $0.9060^*$ {\color{gray}\scriptsize $\pm 0.0513$} & $0.0504^*$ {\color{gray}\scriptsize $\pm 0.0196$} & $26.3458^*$ {\color{gray}\scriptsize $\pm 2.4328$} & $0.8373^*$ {\color{gray}\scriptsize $\pm 0.0772$} & $0.0392^*$ {\color{gray}\scriptsize $\pm 0.0170$} & $28.4788^*$ {\color{gray}\scriptsize $\pm 2.0936$} & $0.8967^*$ {\color{gray}\scriptsize $\pm 0.0776$} \\
\midrule
\multirow{2}{*}{INN (Ours)}
 & \xmark & $0.0471^*$ {\color{gray}\scriptsize $\pm 0.0130$} & $26.8188^*$ {\color{gray}\scriptsize $\pm 2.1457$} & $0.9008^*$ {\color{gray}\scriptsize $\pm 0.0517$} & $0.0509^*$ {\color{gray}\scriptsize $\pm 0.0134$} & $26.1700^*$ {\color{gray}\scriptsize $\pm 2.2764$} & $0.8326^*$ {\color{gray}\scriptsize $\pm 0.0726$} & $0.0391^*$ {\color{gray}\scriptsize $\pm 0.0066$} & $28.2679^*$ {\color{gray}\scriptsize $\pm 1.3878$} & $0.8881^*$ {\color{gray}\scriptsize $\pm 0.0296$} \\
 & \cmark & $\mathbf{0.0455}$ {\color{gray}\scriptsize $\pm 0.0131$} & $\mathbf{27.1434}$ {\color{gray}\scriptsize $\pm 2.2227$} & $\mathbf{0.9095}$ {\color{gray}\scriptsize $\pm 0.0519$} & $\mathbf{0.0467}$ {\color{gray}\scriptsize $\pm 0.0130$} & $\mathbf{26.9338}$ {\color{gray}\scriptsize $\pm 2.3592$} & $\mathbf{0.8521}$ {\color{gray}\scriptsize $\pm 0.0671$} & $\mathbf{0.0346}$ {\color{gray}\scriptsize $\pm 0.0075$} & $\mathbf{29.3818}$ {\color{gray}\scriptsize $\pm 1.6772$} & $\mathbf{0.9120}$ {\color{gray}\scriptsize $\pm 0.0326$} \\
\bottomrule[1.2pt]
\end{tabular}%
}
\end{table*}

\subsection{Ablation Studies}

\subsubsection{Ablation on Loss Functions and Test-Time Adaptation}

We conducted ablation studies to evaluate the contributions of the image reconstruction loss ($\mathcal{L}_{\text{recon}}$), the structure consistency loss ($\mathcal{L}_{\text{structure}}$), and TTA. Quantitative results are summarized in Table~\ref{tab:ablation}.

Starting from the baseline configuration (top row in Table~\ref{tab:ablation}), where the model was trained using only the similarity loss ($\mathcal{L}_{\text{sim}}$) and smoothness loss ($\mathcal{L}_{\text{smooth}}$), progressively adding each component led to consistent improvements across all metrics and datasets. In particular, incorporating $\mathcal{L}_{\text{structure}}$ resulted in pronounced performance gains on both ID and OOD datasets, demonstrating its effectiveness in enforcing anatomically consistent image translation. TTA provided limited improvement on the ID dataset but yielded substantial gains on OOD datasets, highlighting its importance for robustness under distribution shifts. With all components enabled, the full SACRED model achieved the best performance across all metrics and datasets, and significantly outperformed all ablated variants.

\subsubsection{Analysis of Test-Time Adaptation}

We analyzed the effect of the similarity loss function and the number of optimization iterations used during TTA. Specifically, we compared the proposed $\text{LNCC}^2$ loss with the $L_1$ loss, which was used during initial training and evaluated here as an alternative TTA objective. As shown in Fig.~\ref{fig:tta_analysis}, using $\text{LNCC}^2$ consistently outperformed using $L_1$ across all three quantitative metrics, including NRMSE, PSNR, and SSIM. In particular, for SSIM, the $L_1$ loss led to a gradual performance degradation as the number of iterations increased, whereas the $\text{LNCC}^2$ loss maintained stable performance after convergence. In addition, all three metrics were nearly saturated after approximately 500 iterations, as indicated by the gray vertical dashed line in Fig.~\ref{fig:tta_analysis}. Based on this observation, we fixed the number of TTA iterations to 500 in all experiments.

\begin{figure}[!ht]
    \centering
    \includegraphics[width=0.9\linewidth]{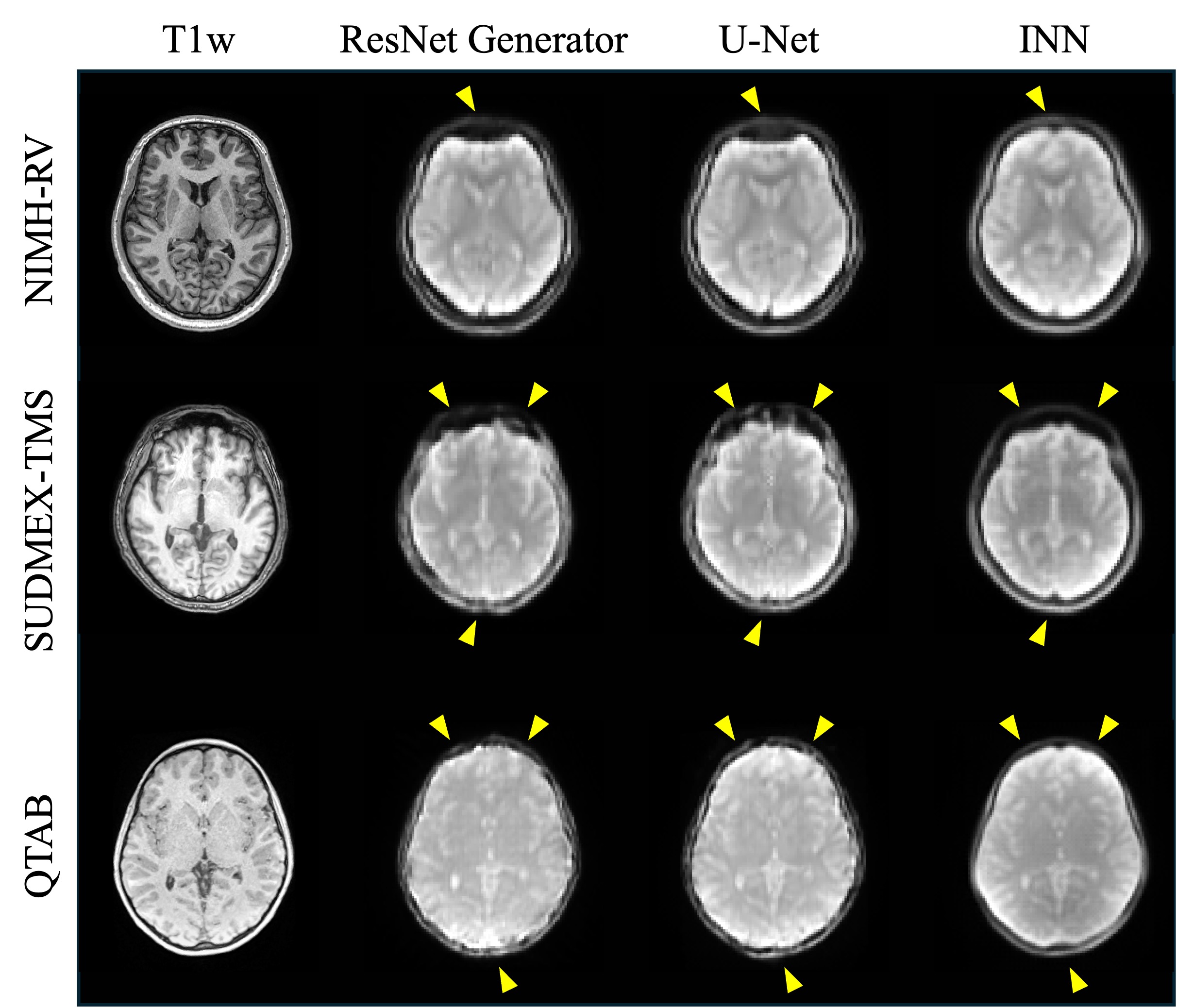}
    \caption{Ablation study on the image translation network architecture. Generated pseudo BOLD images are shown for each architecture across the ID and OOD datasets. Yellow arrowheads indicate regions where the INN-based architecture better preserves anatomical structures from the T1w image than the ResNet- and U-Net-based architectures.}
    \label{fig:pseudo}
\end{figure}

\subsubsection{Ablation on the Image Translation Network Architecture}

We conducted an ablation study on the architecture of the image translation network to evaluate its ability to preserve structural consistency during the image translation process. Specifically, we compared the proposed INN-based architecture with a ResNet-based generator~\cite{zhu_unpaired_2020, johnson_perceptual_2016} and U-Net~\cite{ronneberger2015u}, as these architectures are widely used in medical image translation tasks~\cite{chen2025medical}. For the ResNet- and U-Net-based models, the image reconstruction loss ($\mathcal{L}_{\text{recon}}$) was omitted because these architectures are not invertible. Except for this difference, all other experimental settings were kept the same as those used in the proposed framework.

As shown in Fig.~\ref{fig:pseudo}, the INN-based image translation network better preserved structural consistency between the T1w image and the generated pseudo BOLD image. The improved structural fidelity of the generated pseudo-BOLD images led to better quantitative performance in the distortion-corrected BOLD images, as shown in Table~\ref{tab:ablation_inn}. Moreover, on the OOD datasets, including SUDMEX-TMS and QTAB, the INN-based architecture produced pseudo-BOLD images with clearer brain boundaries, whereas the ResNet- and U-Net-based architectures showed less well-defined boundaries and less consistent anatomical structures, as highlighted by the yellow arrowheads in Fig.~\ref{fig:pseudo}.

\begin{figure}[!ht]
    \centering
    \includegraphics[width=0.9\linewidth]{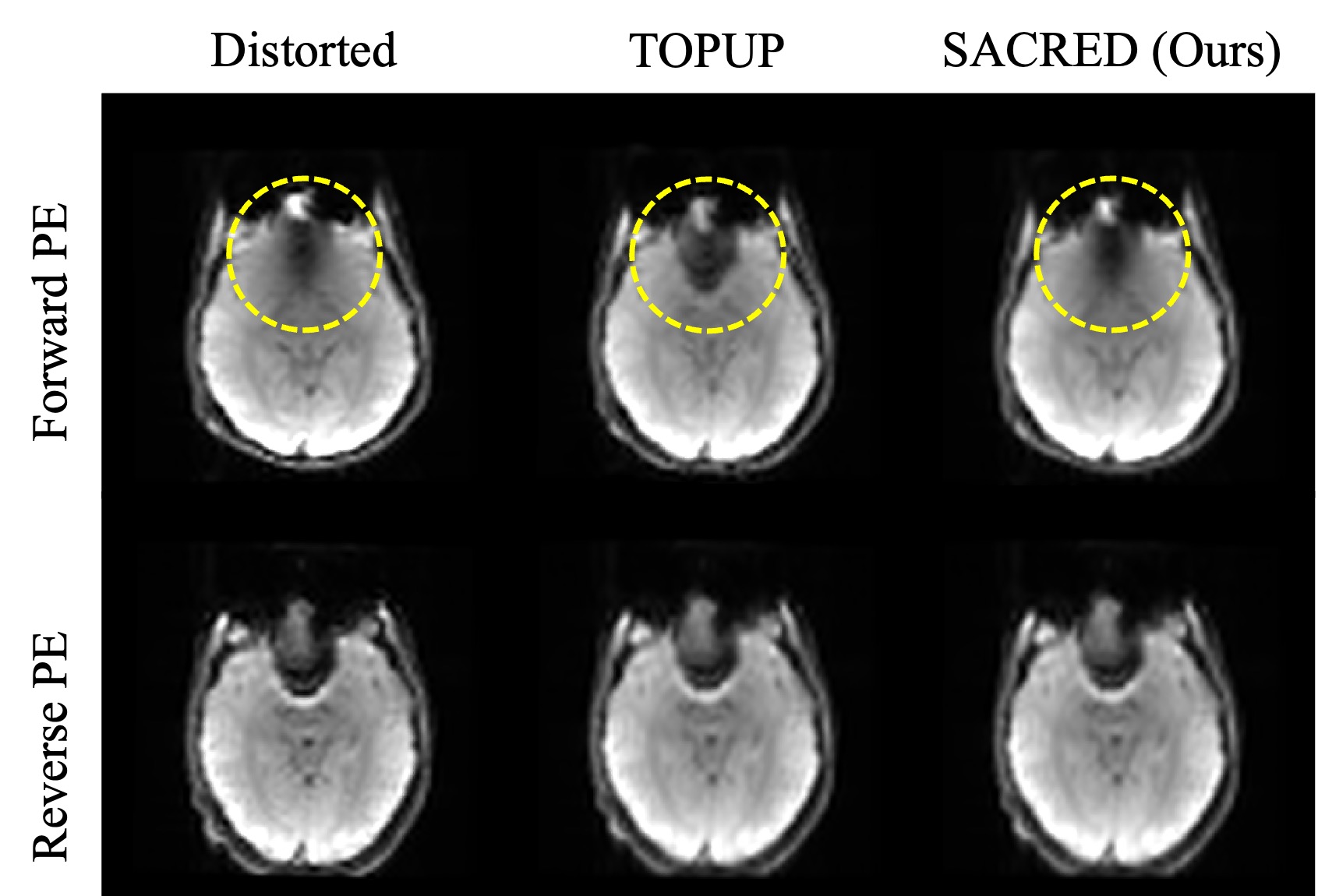}
    \caption{Representative example illustrating a limitation of TOPUP correction in the presence of signal dropout. The first and second rows show forward and reverse PE BOLD images, respectively. Yellow dashed circles in the forward PE row indicate the region where TOPUP correction shows aggravated signal dropout.}
    \label{fig:sacredvstopup}
\end{figure}

\section{Discussion and Conclusion}

In this study, we proposed SACRED, a calibration scan-free fMRI SDC framework that requires only a routinely acquired T1w image and a unidirectional PE BOLD image. By integrating an image translation network with an image registration network, the proposed framework can be optimized in an unsupervised manner using a mono-contrast similarity objective, without requiring ground-truth TOPUP-corrected images.

The INN-based bijective mapping, together with the reconstruction constraint enabled by its inherent invertibility and the structure consistency loss based on MIND-SSC, contributed to improved preservation of anatomical consistency during the image translation process. In addition, TTA, which can be efficiently performed at inference time, further enhanced distortion correction performance, particularly on OOD datasets.

Experimental results demonstrated that SACRED consistently outperformed existing fMRI SDC methods on both ID and OOD datasets. While several learning-based approaches exhibited degraded performance under distribution shifts, SACRED showed improved robustness to both scanner shifts and population (age) shifts. The inferior performance of Synth may be attributed to its correction strategy, which permits deformations along all spatial directions (x, y, and z), potentially leading to over-alignment with the anatomical image. In contrast, most existing correction methods, including SACRED, restrict distortion correction to the PE direction, consistent with prior knowledge of susceptibility-induced distortions in EPI.

Compared with SyN-SDC, a conventional non-learning-based method, SACRED showed relatively smaller performance gains on OOD datasets than on the ID dataset. This behavior may be explained by the data-scaling characteristics of learning-based methods, whose performance typically improves with increased training data diversity and scale. Importantly, because SACRED does not rely on distortion-corrected ground-truth images, it can be trained using widely available datasets containing only unidirectional PE BOLD images and corresponding T1w images. We therefore expect that training SACRED on larger and more diverse datasets, encompassing a broader range of scanners, acquisition protocols, and subject populations, could further enhance its robustness and performance under distribution shifts.

The main difference between GDCNet and SACRED lies in the use of the image translation process. The substantial performance gap between the two methods indicates that directly aligning distorted BOLD images to anatomical T1w images is limited by the contrast mismatch between BOLD and T1w images. These findings suggest that bridging the contrast gap through image translation is essential for reliable calibration scan-free fMRI distortion correction.

Although TOPUP-corrected BOLD images were used as the ground truth for quantitative evaluation, TOPUP itself may produce suboptimal corrections in regions affected by severe susceptibility gradients or signal dropout. In such cases, metrics computed with respect to the TOPUP-corrected ground truth may penalize plausible corrections produced by SACRED. As shown in Fig.~\ref{fig:sacredvstopup}, TOPUP correction aggravated signal dropout in the highlighted frontal region, whereas SACRED preserved the signal pattern without further exacerbating signal loss. This behavior may be related to the meet-in-the-middle correction strategy of TOPUP, in which forward- and reverse-PE images are jointly corrected toward an intermediate representation. When one PE direction exhibits more severe signal dropout, the corrected image may inherit local imperfections from the affected image. This example highlights that TOPUP-corrected images, while serving as a practical ground truth for quantitative evaluation, may still contain local imperfections in regions with severe signal dropout.

One practical consideration of the proposed approach is the additional computational cost associated with TTA. In this study, we empirically fixed the number of TTA iterations to 500 for all test subjects, resulting in inference times on the order of tens of seconds on a GPU. Although this fixed-iteration strategy provided consistent performance improvements, it may not be optimal for all datasets or subjects. Improving computational efficiency through adaptive strategies, such as dataset- or subject-specific early stopping criteria, could better balance correction accuracy and inference cost.

An interesting direction for future work would be to investigate the impact of distortion correction on downstream fMRI analyses, such as functional connectivity. Since all methods, including the TOPUP reference, estimate a single VDM for each BOLD run from the time-averaged 3D image derived from the corresponding motion-corrected 4D BOLD data and apply it uniformly across all time points, the relative performance is expected to be largely consistent with the image-based evaluation presented here.

The proposed EPI SDC framework may be extended to other imaging techniques that employ EPI acquisitions, including diffusion-weighted imaging, arterial spin labeling, and chemical exchange saturation transfer imaging. Evaluating the framework across these applications will further establish its broader applicability.

\bibliographystyle{IEEEtran}
\bibliography{ref}

\end{document}